\documentclass{article}
\usepackage{spconf,amsmath,amssymb,graphicx,hyperref,xcolor,booktabs}
\hypersetup{hidelinks}

\def\x{{\mathbf x}}
\def\e{{\mathbf e}}
\def\z{{\mathbf z}}

\title{HOW MUCH AUDIO IS LEFT IN AN EMBEDDING?\\
AN INVERSION AUDIT OF AUDIO ENCODERS\\[4pt]}

\name{Marios Glytsos \qquad Brian McFee}
\address{Music and Audio Research Laboratory, New York University, NY, USA}

\begin{document}
% \ninept
\maketitle

\begin{abstract}
Pretrained audio encoders are reused for downstream tasks that are often unknown when the encoder is trained, so their usefulness depends partly on which signal properties survive the pretext objective. We study this retained information through paired source reconstruction. Using a shared Stable Audio Open latent diffusion decoder, we reconstruct five-second, 44.1-kHz stereo music from frozen representations produced by supervised classifiers (VGGish, ConvNeXt), an audio-text contrastive model (CLAP), and a waveform-reconstruction model (EnCodec). These objectives impose different pressures to preserve source detail, while their exposed interfaces vary substantially in temporal and spectral resolution. Evaluating on the Million Song Dataset (MSD), we find clear differences in reconstructability across encoder families, while within encoder comparisons show improved recovery when finer temporal or spectral structure is exposed. Even compressed task oriented embeddings support reconstructions that preserve measurable source specificity and high level musical content.
\end{abstract}


\section{Introduction}
\label{sec:intro}

Pretrained audio encoders increasingly serve as reusable interfaces between audio and downstream systems. Unlike traditional engineered features, learned embeddings range from highly compressed task oriented vectors to dense representations trained for reconstruction, making it unclear how much source information survives these different abstractions. This matters when representations are reused for tasks unknown during training, and when embeddings are shared instead of audio in privacy or copyright sensitive settings. We therefore ask: how much source information, beyond what the training objective required, survives in the representation?

We study this through paired source reconstruction. Given a frozen encoder $\Phi$ and representation $\e=\Phi(\x)$, we train a decoder to reconstruct the paired source $\x$. Because $\Phi$ need not be injective, exact deterministic inversion is not always well defined. A conditional generative decoder is therefore natural. It can use the embedding for source structure that is preserved while relying on its learned audio prior for ambiguous details. Successful reconstruction provides evidence that source specific information is accessible from the embedding. Failure however, may also reflect limitations
of the decoder.

We use a standardized reconstruction audit rather than an encoder specific worst case attack. Adaptive inversion can exploit the structure of each target encoder~\cite{mahendran2015understanding,pizzi2022model}, potentially recovering more information but confounding comparisons across representations. Instead, we start from a strong pretrained generator designed to synthesize audio from highly abstract text conditioning, providing an expressive prior for probing what source information an embedding can support without claiming maximal recoverability.

We compare four representative encoders spanning three training objectives. VGGish~\cite{hershey2017cnn} provides a long-standing classification baseline, ConvNeXt~\cite{pellegrini2023convnext} represents a recent AudioSet trained classifier used as a general purpose audio representation, CLAP~\cite{wu2023laionclap} is a widely used audio/text representation, and EnCodec~\cite{defossez2022encodec} provides a reconstruction trained contrast. Their objectives impose different requirements on source detail. Classification rewards discriminative information, contrastive learning emphasizes language aligned semantics, and reconstruction requires local signal detail. 

% For CLAP and ConvNeXt, we additionally vary the exposed temporal or spectral structure to examine how interface compression affects reconstructability independently of the training objective.

We focus on music because reconstructing a recording requires preserving fine grained pitch, timbre, rhythm, and temporal structure, making it a demanding test of how much source information survives an embedding. Music is also a practically relevant setting because licensing constraints often motivate sharing derived representations rather than the original recordings, and instrumental music already lies within the pretrained decoder's domain, reducing domain shift.

The paper makes the following contributions: (1) we adapt a pretrained latent diffusion audio generator to paired source reconstruction from frozen embeddings, and (2) we compare classification, contrastive, and reconstruction trained encoders. The comparison quantifies reconstructability and provides caution when embeddings derived from sensitive recordings are distributed in place of audio.

Code, listening examples and more detailed per-genre scores are available at {\scriptsize\url{https://mariosgly.github.io/research/audio-embedding-inversion/}.}

\section{Related Work}
\label{sec:related}

Representation inversion has long been used to study what information learned features retain. In vision, Mahendran and Vedaldi optimize inputs to match target representations~\cite{mahendran2015understanding}, while Dosovitskiy and Brox learn a decoder that reconstructs image structure from high-level features ~\cite{dosovitskiy2016inverting}. Related work instead frames inversion as a privacy attack~\cite{fredrikson2015model}. 

In audio, Comanducci et al.~\cite{comanducci2021reconstructing} reconstruct speech from intermediate CNN features, while Polyak et al.
~\cite{polyak2021speech} study resynthesis from discrete self-supervised
representations and preservation of content, prosody, and speaker identity. More recent attacks recover audio from intermediate activations ~\cite{zhang2025invercrs} or shared gradients~\cite{zeng2025gradient}. Our goal differs in comparing paired source reconstructability across independently trained representation families using a common pretrained decoder architecture, separately finetuned for each interface.

Audio representations also increasingly serve as conditioning signals for generative models. AudioLDM~\cite{liu2023audioldm}, AudioLDM 2
~\cite{liu2024audioldm2}, and Diff-A-Riff~\cite{nistal2024diffariff} show that learned representations can guide realistic, condition-compatible synthesis. These objectives, however, do not require recovery of the particular recording that produced the condition. 

A complementary line of work studies representation sensitivity to controlled transformations. Kim et al.~\cite{kim2019nearby}, Srivastava et al.~\cite{srivastava2022robustness}, and Deng et al.~\cite{deng2025investigating} examine how known audio changes affect embedding geometry. These methods test which variations are preserved or suppressed, reconstruction instead asks what source-specific information remains jointly recoverable. We apply this view across independently trained representations using a standardized generative probe.

\section{Embedding Reconstruction Audit}
\label{sec:method}

\subsection{Problem formulation}

Let $\x$ denote an audio waveform, $\Phi$ a frozen encoder, and
$\e=\Phi(\x)$ its representation. For each encoder, we train a separate
conditional generator $G_{\Phi}(\e,\epsilon\mid\theta)$, where $\theta$
denotes the generator parameters and $\epsilon\sim\mathcal{N}(0,I)$ the
initial diffusion noise. Training uses paired audio and reconstruction is evaluated on held out tracks. All conditions share the same pretrained initialization, generator architecture, target VAE, and training protocol, only the conditioning interface changes. 

\subsection{Exposed interfaces}

Although all targets are fixed five second crops, the encoder interfaces differ substantially in temporal and spectral resolution. We reconstruct fixed five second crops in every condition. Table~\ref{tab:representations} therefore reports their temporal granularity and the resulting conditioning shape for a five-second crop. We additionally construct alternative CLAP and ConvNeXt interfaces to isolate the effect of temporal and spectral compression while keeping the underlying encoder fixed.

\begin{table}[t]
\centering
\resizebox{\columnwidth}{!}{%
\begin{tabular}{lcc}
\toprule
\textbf{Condition} & \textbf{Temporal granularity} & \textbf{Input shape} \\
\midrule
VGGish & 0.96 s & $128\times(6\text{--}7)$ \\
CLAP 5s & 5 s & $512\times1$ \\
CLAP 1s & 1 s & $512\times5$ \\
ConvNeXt pooled & $\approx0.32$ s & $768\times(16\text{--}17)$ \\
ConvNeXt flattened & $\approx0.32$ s & $5376\times(16\text{--}17)$ \\
EnCodec $z_q$ & 6.7 ms & $128\times750$ \\
\bottomrule
\end{tabular}%
}
\caption{Frozen interfaces exposed for one 5 second target. Shapes are feature dimension $\times$ number of conditioning tokens.}
\label{tab:representations}
\end{table}

VGGish processes mono 16-kHz audio into non-overlapping 0.96-s, 64 bin
log-mel patches, producing one 128 dimensional embedding per patch. We use the
released PCA-whitened 8-bit output, rescaled from $[0,255]$ to $[-1,1]$.

LAION-CLAP~\cite{wu2023laionclap} processes mono 48-kHz audio with an
HTSAT base tower and produces a 512-dimensional embedding. \emph{CLAP 5s}
encodes the full crop once, while \emph{CLAP 1s} independently encodes five
one second segments.

AudioSet ConvNeXt Tiny~\cite{pellegrini2023convnext} processes mono 32-kHz
audio and produces a final $T\times768\times7$ feature map, yielding 16--17
temporal positions per crop. The \emph{pooled} interface averages the seven
frequency positions, while the \emph{flattened} interface preserves them as
5376-dimensional tokens. The original work similarly retains 31 temporal
embeddings over ten seconds for audio captioning.

EnCodec~\cite{defossez2022encodec} processes 48-kHz stereo audio and produces
the quantized representation $z_q\in\mathbb{R}^{T\times128}$ at 150 frames/s,
yielding 750 tokens per crop. Our probe conditions only on $z_q$, omitting the
native scale value. EnCodec serves as a reconstruction trained positive control.

\begin{figure*}[t]
    \centering
    \includegraphics[width=1.0\linewidth]{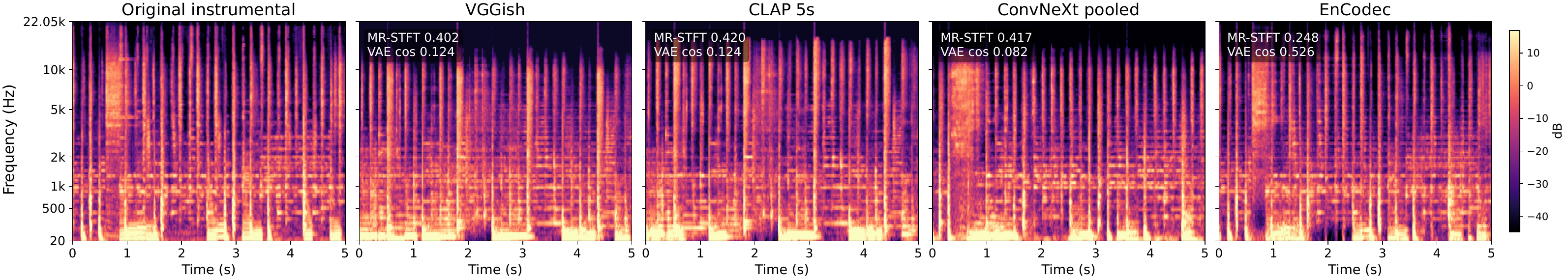}
    \caption{Qualitative reconstruction of an MTG-Jamendo track using the standard interfaces.}
    \label{fig:spectrogram_comparison}
\end{figure*}

\subsection{Conditional latent diffusion probe}

We initialize each decoder from Stable Audio Open variant 2.0 trained on Creative Commons data~\cite{evans2024stableaudioopen}. Its fixed variational autoencoder maps five second, 44.1kHz stereo audio to 64 channel latent representations with a downsampling ratio of 2048. The pretrained 24 layer diffusion transformer has width 1536 and 24 attention heads and cross attends to 768 dimensional tokens. The latent target, autoencoder, and transformer architecture are identical across encoders.

We replace the pretrained T5 conditioning interface with each frozen audio
representation. Each crop yields a sequence $\e\in\mathbb{R}^{T\times D}$ whose length and dimension depend on the encoder. A learned linear layer maps $D$ to the transformer's 768 dimensional conditioning space. The pretrained cross attention was learned for T5 features, while the replacement audio features differ in scale, geometry, and temporal density. We therefore initialize the new projection close to zero, with $W_{\mathrm{proj}}\sim\mathcal{N}(0,10^{-3})$ and $b_{\mathrm{proj}}=0$, to reduce the initial perturbation introduced by the new conditioning interface. Our experiments indicate more stable optimization and better convergence with this initialization.

Dense conditioning also makes temporal correspondence unnecessarily difficult  if every latent query can attend equally to every conditioning token. We assign normalized times $t_i,s_j\in[0,1]$ to latent and conditioning tokens  and modify the cross-attention score as 
\begin{equation} 
  a_{ij}=\frac{\mathbf q_i^\top\mathbf k_j}{\sqrt d} 
  +b_h\!\left(\operatorname{bucket}(|t_i-s_j|)\right). 
\end{equation} 
The learned bias is head-specific and initialized to zero. A hard mask also restricts attention to $|t_i-s_j|\leq0.2$. Together, these terms express the weak prior that corresponding audio regions should remain locally aligned. A query reconstructing 5.0--5.2\,s of audio is therefore biased toward conditioning tokens from the same region rather than distant context. This prior also eases finetuning under our limited compute budget, for interfaces with many conditioning tokens, such as EnCodec. The bucket widths and mask radius were set heuristically based on temporal locality, while learning rate schedules were selected using validation-set performance.

\subsection{Training objective}

Stable Audio Open uses the $v$-prediction objective. Let $\z_0$ denote the fixed VAE representation of the target audio and $\epsilon\sim\mathcal{N}(0,I)$. At diffusion time $\tau$,

\begin{equation}
\z_\tau=\alpha_\tau\z_0+\sigma_\tau\epsilon,\qquad
\mathbf v_\tau=\alpha_\tau\epsilon-\sigma_\tau\z_0,
\end{equation}
where $\alpha_\tau$ and $\sigma_\tau$ are the signal and noise coefficients with $\alpha_\tau^2+\sigma_\tau^2=1$. The diffusion transformer $f_\theta$ predicts $\mathbf v_\tau$ conditioned on the encoder representation $\e$, and is trained to minimize 

\begin{equation}
\mathcal{L}_{v}=
\mathbb{E}_{\x,\tau,\epsilon}
\left\|f_\theta(\z_\tau,\tau,\e)-\mathbf v_\tau\right\|_2^2.
\end{equation}

Because this objective does not directly optimize source-waveform similarity, validation loss is used as an optimization diagnostic rather than a measure of reconstruction quality.

Classifier free guidance (CFG) normally trades off conditional and unconditional predictions by randomly dropping the condition during training~\cite{ho2022classifierfreediffusionguidance}. This is unnecessary for our setting, the goal is to reconstruct from the observed embedding rather than trade conditioning fidelity for sample diversity. We therefore use zero conditioning dropout and guidance weight $w=1$, retaining only the conditional prediction.

At inference, we use a 150-step deterministic DDIM sampler ($\eta=0$).
Randomness therefore enters only through the initial noise realization, no  noise is injected during denoising. This reduces sampling variability that would otherwise confound comparisons between paired reconstructions.

% \begin{table*}[t]
% \centering
% \small
% \setlength{\tabcolsep}{5pt}
% \begin{tabular}{lccccc}
% \toprule
% Encoder &
% MR-STFT $\downarrow$ &
% SAO-VAE cos. $\uparrow$ &
% R@1 $\uparrow$ &
% R@10 $\uparrow$ &
% MusicNN Top-10 $\uparrow$ \\
% \midrule

% VGGish
% & 0.21 $\pm$ 0.08
% & 0.15 $\pm$ 0.07
% & 0.004
% & 0.021
% & 7.33 $\pm$ 1.44 \\

% CLAP 5s
% & 0.20 $\pm$ 0.08
% & 0.16 $\pm$ 0.07
% & 0.006
% & 0.034
% & 7.51 $\pm$ 1.34 \\

% CLAP 1s
% & 0.19 $\pm$ 0.08
% & 0.18 $\pm$ 0.07
% & 0.024
% & 0.091
% & 7.71 $\pm$ 1.28 \\

% ConvNeXt pooled
% & 0.18 $\pm$ 0.08
% & 0.19 $\pm$ 0.07
% & 0.094
% & 0.201
% & 7.99 $\pm$ 1.21 \\

% ConvNeXt flattened
% & 0.18 $\pm$ 0.07
% & 0.24 $\pm$ 0.07
% & 0.352
% & 0.539
% & 8.12 $\pm$ 1.16 \\

% EnCodec
% & 0.13 $\pm$ 0.06
% & 0.47 $\pm$ 0.16
% & 0.894
% & 0.913
% & 8.47 $\pm$ 1.07 \\

% \midrule

% EnCodec native
% & 0.10 $\pm$ 0.05
% & 0.89 $\pm$ 0.05
% & 1.000
% & 1.000
% & 9.09 $\pm$ 0.79 \\

% \bottomrule
% \end{tabular}
% \caption{Reconstruction on our five second MSD test crops. Values are mean $\pm$ standard deviation where applicable. R@1/R@10 denote paired-source retrieval. EnCodec native uses its original decoder.}
% \label{tab:msd-reconstruction-main}
% \end{table*}

\subsection{Dataset}

We use the Million Song Dataset (MSD)~\cite{bertinmahieux2011msd} with the
Tagtraum CD2 genre annotations~\cite{schreiber2015genre}. After requiring complete paired audio and features, our artist-disjoint split contains 219,810 training, 26,190 validation, and 26,367 test tracks across 15 genres. Because Stable Audio Open was trained on royalty-free instrumental audio and sound effects~\cite{evans2024stableaudioopen}, we remove vocals with Demucs~\cite{defossez2021hybrid} to prevent the base generator's domain limitations from bottlenecking the reconstruction audit, and use the resulting instrumental mixtures for both encoder extraction and reconstruction targets. 

\textbf{Implementation details.}
All models are trained across 4 GPUs with a per device batch size of 32. Finetuning the conditional latent diffusion decoder takes approximately 36 hours per encoder conditioning interface.

\subsection{Evaluation}

The natural first question is whether the paired source itself can be
reconstructed from an embedding. We measure acoustic reconstruction using multi-resolution STFT distance (MR-STFT) and cosine similarity in the fixed Stable Audio Open VAE space. The latter helps distinguish waveform reconstruction error from limitations of the generator representation itself, high VAE similarity with poor MR-STFT indicates that acoustically different signals are represented similarly within the decoder's domain. We calibrate VAE similarity against 200k randomly sampled distinct source--source crop pairs. Figure~\ref{fig:reconstruction_violin} compares this reference distribution with the paired source--reconstruction distributions for each encoder. We also retrieve each reconstruction against all 26,367 held-out source crops, R@1 and R@10 report whether its paired source is among the nearest one or ten sources in VAE space.

Acoustic mismatch need not imply loss of musical content, so we also compare source--reconstruction pairs using the MSD-trained MusicNN ~\cite{pons2019musicnn}. We report their normalized top 10 tag overlap $|\mathcal{T}_{\mathrm{src}}\cap\mathcal{T}_{\mathrm{rec}}|/10$, as a complementary measure of preserved high-level musical content.

\begin{table}[t]
\ninept
\centering
\setlength{\tabcolsep}{3pt}
\begin{tabular}{lccccc}
\toprule
Encoder & STFT$\downarrow$ & VAE$\uparrow$ & R@1$\uparrow$ & R@10$\uparrow$ & Tag$\uparrow$ \\
\midrule

VGGish & .21$\pm$.08 & .15$\pm$.07 & .004 & .021 & .733$\pm$.144 \\
CLAP 5s & .20$\pm$.08 & .16$\pm$.07 & .006 & .034 & .751$\pm$.134 \\
CLAP 1s & .19$\pm$.08 & .18$\pm$.07 & .024 & .091 & .771$\pm$.128 \\
ConvNeXt pool. & .18$\pm$.08 & .19$\pm$.07 & .094 & .201 & .799$\pm$.121 \\
ConvNeXt flat. & .18$\pm$.07 & .24$\pm$.07 & .352 & .539 & .812$\pm$.116 \\
EnCodec & .13$\pm$.06 & .47$\pm$.16 & .894 & .913 & .847$\pm$.107 \\
\midrule
EnCodec native & .10$\pm$.05 & .89$\pm$.05 & 1.000 & 1.000 & .909$\pm$.079 \\

\bottomrule
\end{tabular}
\caption{Reconstruction on our five-second MSD test crops. Values are mean
$\pm$ standard deviation where applicable. R@1/R@10 denote paired-source
retrieval. Tag is normalized MusicNN top-10 tag overlap. EnCodec native uses
its original decoder.}
\label{tab:msd-reconstruction-main}
\end{table}

\section{Results and Discussion}
\label{sec:results}

Table~\ref{tab:msd-reconstruction-main} shows a consistent ordering in reconstructability, from VGGish and CLAP through ConvNeXt to EnCodec. This ordering is consistent with the degree of source detail encouraged by the different pretraining objectives, though the encoder families also differ in architecture and training data. The within encoder comparisons are more direct, finer CLAP temporal resolution and preserving ConvNeXt frequency structure both improve reconstruction, showing that interface compression itself limits recoverability.

EnCodec serves as a positive control for the common probe, while its native decoder provides a topline reference. The common probe approaches native EnCodec in MR-STFT and paired-source retrieval, and listening examples are recognizably close to the source, providing qualitative confirmation that the decoder can effectively use detailed conditioning. At the same time, these reconstructions remain farther away and more variable in SAO-VAE space. 
%showing that MR-STFT and VAE similarity capture different aspects of reconstruction. 
Across different models MR-STFT and SAO-VAE similarity are only weekly correlated ($|r|=0.02$--$0.17$, $R^2<0.03$), which supports their use as complementary rather than interchangeable reconstruction measures capturing different aspects of the reconstruction.  

\begin{figure}[t]
    \centering
    \includegraphics[width=0.95\linewidth]{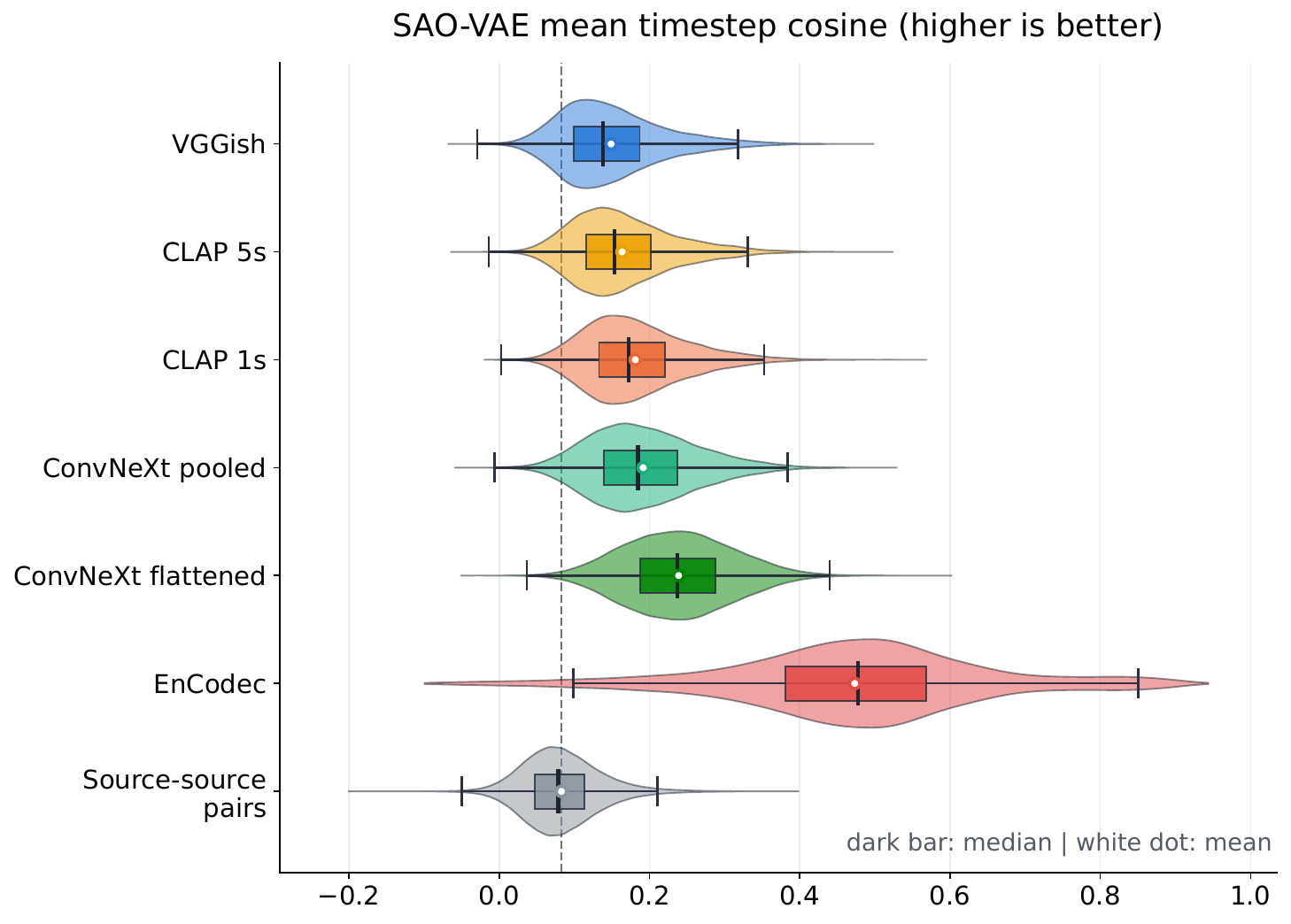}
    \caption{Distribution of paired source--reconstruction SAO-VAE cosine
    similarity across encoder conditions, compared with 200k randomly sampled
    distinct source--source pairs.}
    \label{fig:reconstruction_violin}
\end{figure}

The complementary metrics reveal that musical content and source specificity are not equivalent. For compressed representations such as VGGish and CLAP, reconstructions may miss exact pitch or timing while still retaining similar timbral texture and mixtures of instruments. The relatively strong MusicNN tag agreement is consistent with this behavior. Paired source retrieval, by contrast, improves much more sharply as finer temporal and spectral structure is exposed. Figure~\ref{fig:reconstruction_violin} provides a distributional view of source specificity, most source--reconstruction similarities lie above the random source--source distribution, indicating that the reconstructions remain systematically closer to their paired source than to an unrelated test track.

Figure~\ref{fig:spectrogram_comparison} reveals several qualitative differences between the representations. Notably, CLAP 5s retains frequency-specific temporal patterns despite using a single embedding, with low-frequency regions switching on and off similarly to the source. VGGish likewise preserves coarse rhythm and the overall spectral character, but follows source-specific harmonic structure less faithfully, consistent with its relatively strong MusicNN agreement despite weaker acoustic reconstruction. ConvNeXt pooled more closely tracks both the temporal pattern and harmonic structure, although frequency pooling removes finer spectral detail that the generative decoder must infer. EnCodec most closely reproduces the source's time--frequency structure.

Conditioning ablations which can be found on the project page, show clearer gains in validation denoising loss from near zero projection initialization and temporally local attention than in final reconstruction metrics, particularly for denser interfaces, while learning rate choice also affects reconstruction variability.

\section{Conclusion}
\label{sec:conclusion}

Audio embeddings are useful abstractions, but downstream utility does not reveal how much source detail remains reconstructable. We frame inversion as a controlled audit and compare held-out source recovery from supervised, contrastive, and reconstruction-trained representations using a shared pretrained diffusion probe. Future work should extend this analysis beyond music, to broader encoder families and more direct privacy leakage measures.

\section*{Acknowledgments}
This work was supported in part through the NYU IT High Performance Computing resources, services, and staff expertise.
\begingroup
\ninept
\bibliographystyle{IEEEbib}
\bibliography{strings,references}
\endgroup

\end{document}